\documentclass{aastex631}

\graphicspath{{./}{figures/}}

\begin{document}

\title{Limitations and biases in the retrieval of the polar magnetic field I: the role of the magnetic filling factor in Milne-Eddington inversions of simulated Hinode/SP data}


\correspondingauthor{Rebecca Centeno}
\email{rce@ucar.edu}

\author[0000-0002-1327-1278]{Rebecca Centeno}
\affiliation{High Altitude Observatory (NCAR)\\
3080 Center Green Dr., Boulder, CO, 80301, USA}

\author[0000-0002-0189-5550]{Ivan Mili\'{c}}
\affiliation{Leibniz Institute for Solar Physics (KIS), Sch\"{o}nek Str. 6, 79111 Freiburg, Germany}

\author[0000-0001-5850-3119]{Matthias Rempel}
\affiliation{High Altitude Observatory (NCAR)\\
3080 Center Green Dr., Boulder, CO, 80301, USA}

\author[0000-0001-6119-0221]{Nariaki V. Nitta}
\affiliation{Lockheed Martin Advanced Technology Center, Palo Alto, CA, 94304, USA}

\author[0000-0003-4043-616X]{Xudong Sun}
\affiliation{University of Hawaii}



\begin{abstract}

We study the extent to which Milne-Eddington inversions are able to retrieve and characterize the magnetic landscape of the solar poles from observations by the spectropolarimeter onboard Hinode. In particular, we evaluate whether a variable magnetic filling factor is an adequate modeling technique for retrieving the intrinsic magnetic properties from every pixel in the polar field of view. We first generate synthetic spectra emerging from a numerical simulation of a "plage" region at an inclined line of sight of 65$^{\circ}$, and degrade the data to emulate real observations. Then, we invert the synthetic spectra with two Milne-Eddington inversion codes that feature different treatments of the magnetic filling factor, and relate the retrieved magnetic quantities back to their original values in the simulation cube. We find that while the {\em apparent} retrieved magnetic properties map well the spatially-degraded simulation, the {\em intrinsic} magnetic quantities bear little relation to the magnetic field at the native resolution of the simulation. We discuss the systematic biases caused by line-of-sight foreshortening, spatial degradation, photon noise and modeling assumptions embedded in the inversion algorithm.

\end{abstract}

\section{Introduction} \label{sec:intro}

The magnetic nature of the polar caps of the Sun has profound implications for the solar cycle, space weather and space climate. Mostly unipolar in character, the solar poles provide the bulk of the open magnetic flux permeating the heliosphere. They are also the sources of the fast solar wind and they are believed to provide the magnetic seed for the next solar cycle \citep{charbonneau, petrie2017, nandy}.
The fields of the Sun's poles have been observed routinely since the 1960s through line-of-sight (LOS) magnetograms. These observations revealed the cyclic behavior of the poles in the context of the butterfly diagram and the timing of the reversal of the polar field, both of which are pillars for Babcock's phenomenological description of the solar cycle \citep{babcock1961}.

The radial flux at the poles of the Sun is one of the best predictors of the amplitude of the next solar cycle \citep{dikpati, 2020LRSP...17....2P}. Historically, most observations of magnetic field at the poles are in the form of LOS magnetograms, which only allow the measurement of the magnetic flux along the observer's viewing angle. The radial flux is then reconstructed from the latter using the radial field approximation, which assumes that magnetic fields are always vertical to the local solar surface. \cite{gosain} showed that this approximation leads to systematic errors in the radial field inference that are more pronounced the higher the latitude.

Observations of the full Stokes vector have, in principle, all the information needed to reconstruct the three components of the magnetic field. However, these measurements, along with the field inferences, are not free of challenges. The characterization of the magnetic field in the photosphere of the Sun relies on the interpretation of polarization signals due to the Zeeman effect. There are many challenges associated to the inferences of weaker fields, such as those due to the difference in the sensitivity of the linear and circular polarization to the magnetic field, the effects of noise in the inversions, the cancellation of Zeeman-induced polarization signals due to finite spatial resolution, the limitations imposed by spectral and temporal sampling, etc  \citep[see extensive review by ][]{bellot_rubio_LRSP}. At the poles, these challenges are further compounded by severe foreshortening and projection effects along with strong intensity gradients close to the limb \citep{petrieLRSP}. Finally, the inference of the magnetic field vector always yields two mathematically valid solutions which, at the pole, correspond to two physically distinct orientations of the field. This ambiguity cannot be resolved without additional physical assumptions \citep{metcalf2006, leka2009}.

\cite{tsuneta2008} use data from the spectropolarimeter \citep[SP,][]{hinodeSP} onboard the Hinode \citep{hinode} spacecraft to analyze the vector magnetic field landscape of the solar poles at unprecedented high-spatial resolution. Spectral line inversions using the MILne–Eddington inversion of the pOlarized Spectra \citep[MILOS,][]{MILOS} code reveal large unipolar nearly-vertical patches of kG fields mixed with smaller magnetic structures that are more horizontal in nature. The polarity of the large vertical patches coincides with the dominant magnetic polarity at the pole. The authors also provide an estimate of the total magnetic flux of the polar cap, and find that it is just shy of that of a typical active region. 
Subsequent works take the analysis of Hinode/SP data further and study the differences between the polar fields and the quiet Sun \citep{Ito2010}, and the evolution of the magnetic landscape of the poles through a polarity reversal \citep{shiota}. However, the precise quantitative estimates of the polar flux reported in these works are contingent on the treatment and assumptions of modeling the magnetic filling factor.

The magnetic filling factor is a common modeling parameter that represents the fraction of the resolution element filled with magnetic field \citep{Jefferies1989}. There is plenty of observational evidence that the magnetic fields on the Sun arrange themselves at spatial scales smaller than typical diffraction limits \citep[e.g.][]{sanchezalmeida2011}. Numerical simulations also support this picture \citep[e.g.][]{matthias2014}. As \cite{landi_book} point out, many of the methods developed for solar magnetometry are based on the simple assumption that the magnetic field is uniform over the spatial extent of a pixel, but this often fails to produce good spectral fits to the full Stokes vector in observations at moderate spatial resolution \citep{Lites1990}. A slightly more complex approach to the spectral line modeling assumes that, even though the magnetic field is still horizontally uniform, it only occupies a fraction $\alpha$ of the pixel's area. Here $\alpha$ is known as the magnetic filling factor and it is often a free model parameter in spectral line inversions. $\alpha$ is allowed to vary between 0 and 1, with $\alpha=1$ implying a fully and uniformly magnetized pixel, and $\alpha=0$ representing a non-magnetized atmosphere.

\cite{xudong2021} study the effect of a variable magnetic filling factor on the inversion results of the Helioseismic and Magnetic Imager \citep[HMI,][]{HMI} on board the Solar Dynamics Observatory \citep{SDO}. Due to the limited spectral resolution, the inversion in the official HMI data pipeline forces the magnetic filling factor to be unity \citep{HMIinversion}, this is, assumes a uniformly magnetized pixel. According to \cite{xudong2021}, this results in magnetic fields that are systematically more inclined towards the plane of the sky than the local radial direction, and generally tilted towards the pole. When repeating the inversions allowing the filling factor to be a free parameter of the inversion model, they find that the inferred fields become more radial, albeit the bias persists. A related work by \cite{ana2021} shows that using the filling factor in HMI inversions also helps mitigate the anomalous hemispheric bias found in plage regions.
In a detailed comparison between HMI and Hinode/SP data and their official inversion schemes, \cite{sainz_dalda2017} finds that the treatment of the magnetic filling factor (assumed to be unity for HMI  and variable for  Hinode/SP) is the principal cause of the differences in the inferred magnetic properties inside plage regions, followed closely by differences in spatial and spectral resolutions.
\cite{pevtsov2021}, on the other hand, argue that the effects of using a variable magnetic filling factor for the inversions of weak polarization signals may result in systematic biases in the inversion results that lead to erroneous interpretation of the magnetic landscape in certain regions of the Sun, the poles being especially vulnerable to these errors.

While these may seem highly-specialized details that only matter to spectral line inversion experts, the impacts of erroneous inferences cascade all the way down to Space Weather prediction. Models of the coronal and heliospheric magnetic structure, which rely on the extrapolation of photospheric magnetograms, systematically underestimate the interplanetary magnetic field that is measured in-situ at 1~AU \citep{linker2017, linker}. The higher solar latitudes play a disproportionate role here since the open magnetic flux  is believed to emanate mainly from the poles of the Sun.

In this work we study the ability of inversion schemes with variable magnetic filling factor to tease out a faithful description of the intrinsic magnetic field from solar polar observations with the Hinode/SP instrument.
In order to evaluate how accurate our inferences of polar magnetic fields are we need to compare the inversion results to a known ground-truth. Here we take the approach of generating synthetic Hinode/SP-like spectra (along a polar line-of-sight) from a radiative magnetohydrodynamic (rMHD) simulation of the Sun's photosphere. Then, we invert the synthetic spectra with two commonly used spectral line inversion codes and we compare the inversion results to the physical quantities in the simulation cube.
After a brief introduction to the formulation of the magnetic filling factor, section \ref{sec:synthetic_spectra} describes the generation of Hinode/SP-like synthetic spectra.
In section \ref{sec:inversion_setup}, we invert the spectra with two Milne-Eddington inversion codes (MILOS and MERLIN). Inter-comparison of the inversion results obtained by the two codes and evaluation of the results against the ground-truth of the simulation are shown in section \ref{sec:inversion_results}. Lastly, we discuss the implications of the results for estimates of the open magnetic flux and close with an outlook into the future.


\section{Formulation of the magnetic filling factor: apparent vs. intrinsic magnetic quantities}

The formulation of the magnetic filling factor $\alpha$ in the model atmosphere stems from the hypothesis that the Stokes vector recorded at each individual resolution element has two contributions: one magnetic (M) and one non-magnetic (NM). The observed Stokes profiles can then be computed as the weighted sum of these two components, where only the magnetic component contributes to the emergent polarization \citep[see, e.g.][]{Jefferies1989, bellot_rubio_LRSP}:

\begin{eqnarray}
\label{eq:filling_factor1}
    I &=& \alpha I_{\rm M} + (1-\alpha)I_{\rm NM} \\
    \label{eq:filling_factor2}
    Q &=& \alpha Q_{\rm M} \\
    \label{eq:filling_factor3}
    U &=& \alpha U_{\rm M} \\
    \label{eq:filling_factor4}
    V &=& \alpha V_{\rm M}  
\end{eqnarray}

This model inherently assumes that the magnetic component uniformly occupies a fraction $\alpha$ of the pixel. The non-magnetic component, which only produces an intensity profile, $I_{\rm NM}$, can be modeled in different ways. It is customary to prescribe this non-magnetic intensity profile as the (local or global) average intensity spectrum across the spatial domain of the observation. In other words, the non-magnetic component is often treated as a stray light spectral profile with a ($1-\alpha$) contribution to the emergent spectrum. Section \ref{sec:inversion_setup} delves into how the stray light is computed in two different data pipelines for Hinode/SP observations.

Under the weak field approximation, one can derive a relationship between the unresolved (or {\em apparent}) and resolved (or {\em intrinsic}) vector-magnetograph measurements \citep{landi_book}.
The latter are the "true" components of the magnetic field vector, typically represented by its spherical coordinates in the line-of-sight reference system: the magnetic field strength, $B$, its inclination with respect to the LOS, $\theta$, and its azimuth in the plane of the sky, $\chi$. 
The {\em apparent} magnetic quantities, on the other hand, are those that would lead to the same (or similar) Stokes profiles if the pixel were fully and uniformly magnetized \citep{Lites2008}. In other words, it is the magnetic field that would be inferred if we assume that $\alpha=1$. They represent the "pixel-averaged" magnetic field components, and are related to the intrinsic ones in the following way \citep{landi_book}:

\begin{eqnarray}
    \label{eq:blos_app}
    B_{\rm LOS}^{\rm app} &=& \alpha B_{LOS} \\
    \label{eq:bt_app}
    B_{\rm T}^{\rm app} &=& \sqrt{\alpha} B_{\rm T} \\
    \label{eq:theta_app}
    {\rm tan}{\theta^{\rm app}} &=& {\rm tan}\big(\frac{\theta}{\sqrt{\alpha}}\big)
\end{eqnarray}

where $B_{\rm LOS} = B cos(\theta)$ is the line-of-sight component of the magnetic field, $B_{\rm T} = B sin(\theta)$ is its transverse component, and the super-index "app" is used to label the {\em apparent} magnetic quantities. $B_{\rm LOS}^{\rm app}$ and $B_T^{\rm app}$ are often referred to as the LOS and transverse {\em apparent} flux densities and even though they have the same physical units as the intrinsic magnetic quantities, hereonafter we use "Mx/cm$^{2}$" for the former and "Gauss" for the latter, following \cite{Lites2008}. The apparent and the intrinsic azimuths are identical ($\chi^{app} = \chi$). Eq. \ref{eq:theta_app} readily shows how the assumption of $\alpha=1$ results in apparent inclination angles that are more transverse to the LOS than they would otherwise be \citep{sanchezalmeida2011}.

Even though Eqs. \ref{eq:blos_app}--\ref{eq:theta_app} are strictly only valid in the weak field regime, in Section \ref{sec:inversion_results} we use the {\em apparent} magnetic quantities to compare the Milne-Eddington inversion results to the spatially-degraded magnetic fields from the simulation. This is justified by the predominantly weak nature of the fields in this MURaM cube and, as the reader will see in due course, the apparent inverted quantities are able to capture the unresolved magnetic properties of the simulation rather well. 

\section{Synthetic spectra}\label{sec:synthetic_spectra}

In this section we describe the generation of synthetic Hinode/SP-like Stokes spectra.
Hinode/SP observes the photospheric Fe {\sc i} line pair at 6301.5 and 6302.5 \AA\ with a nominal spatial resolution of 0.32" and a spectral sampling of 21.5~m\AA.
 
We start from a MURaM simulation of a plage region with a mean vertical magnetic field imbalance of 30~G on any given horizontal cross-section. This simulations is based on the small-scale dynamo simulation 'O16bM' from \citet{matthias2014}. We added a mean vertical magnetic field of 30~G  and evolved the simulation for another 6 hours of solar time to allow for the formation of a magnetic network structure. The last 15 minutes of the simulation were performed with non-gray radiative transfer. The original simulation cube has a sampling of $Dx=Dy=Dz=16$~km in all three spatial directions, with a 24.5~Mm extent in both horizontal dimensions and 7.7~Mm in the vertical direction, $z$ (of which $\sim 6.2$~Mm lie below the base of the photosphere).

We synthesize the Fe {\sc i} spectra along a slanted observing geometry to simulate a polar viewpoint. Specifically, we assume a heliocentric observing angle of $\theta_{\rm LOS}=65^{\circ}$, which corresponds to a low-latitude polar observation with moderate foreshortening ($\sim 870$ arcsec from disk center). To synthesize the spectra emitted from this polar observing geometry, we shear and interpolate the simulation cube to transform the slanted observing LOS into a new vertical direction on a regularly-spaced cartesian grid. To do this, we displace each horizontal layer of the cube by $n \times Dx / tan(\theta_{\rm LOS}) = n \times 34.31$~km, where $34.31$~km corresponds to the horizontal displacement between two consecutive layers for this viewing angle and $n$ is the (integer) layer number starting from the top. The sinusoidal boundary condition of the simulation allows us to wrap around the displaced edges of the simulation cube resulting in no loss of spatial extent in the sheared simulation box. The sheared layers are then interpolated onto a regular cartesian grid with a $16$~km spacing in the horizontal directions. The new vertical distance between layers is $\Delta z = Dz /cos(\theta_{\rm LOS}) = 37.85$~km. Lastly, the vector quantities (magnetic field and velocity) of the original simulation box are projected onto the new cartesian axes to obtain the velocity and the three components of the vector magnetic field in the slanted (LOS) reference frame. 

After this transformation, the slanted LOS in the original simulation box becomes the vertical direction in the sheared cube. We can now calculate the emergent Stokes spectra along the vertical direction in the sheared cube.

The spectral synthesis is carried out with the SIR (Stokes Inversion based on Response functions \cite{SIR}) code.
SIR is a package for the synthesis and inversion of spectral lines formed in the presence of magnetic fields in the Local Thermodynamical Equilibrium (LTE) regime. The code takes into account the Zeeman-induced polarization of the light and deals with all four Stokes parameters of any electric dipole transition and atomic species. 
In synthesis mode, the program calculates the Stokes
spectra emerging from any vertically stratified model atmosphere (where temperature, magnetic field vector, line of sight velocity, as well as the micro- and macro-turbulent velocities are specified  --- although the latter two are not used for the generation of synthetic data in this work). The code solves numerically the radiative transfer equation (RTE) for polarized light to generate the emergent Stokes spectra in a 1D plane-parallel geometry. 

In order to synthesize the Fe {\sc i} 6301 and 6302 \AA\ lines, we take each vertical column in the sheared cube and put it into a 1-D model atmosphere as a function of optical depth in the SIR format to obtain the emergent Stokes profiles. There are $1536 \times 1536 \sim 2.34$ million 1D model atmospheres which result in as many sets of Stokes spectra for the 6301-6302 \AA\ line pair.

\subsection{Degradation of the synthetic spectra}

Once the synthetic spectra are computed, we degrade them to the resolution of Hinode/SP observations. We first carry out the spectral degradation. This is modeled after \cite{hinodeSP}, which found that the Hinode/SP instrument has a gaussian spectral point spread function (PSF) with a full width half maximum (FWHM) of 24.5 m\AA\ and a spectral sampling of 21.5 m\AA. We convolved each spectrum with this gaussian function and rebinned the output to the specified sampling.

After spectral degradation, we spatially foreshorten the pixels\footnote{\url{https://scipy-cookbook.readthedocs.io/items/Rebinning.html}} along one of the horizontal dimensions (coincident with the shearing direction) to mimic the projection effect due to an inclined LOS. The foreshortened cube has horizontal dimensions of $1536 \times 649$ pixels, and the geometric distance between pixels along the foreshortened direction is now $\Delta x_f = \Delta_x / \mu = 37.86$~km (where $\mu = cos(\theta_{\rm LOS})$ is the cosine of the heliocentric angle).

\begin{figure}[ht!]
\centering
\includegraphics[width=1.0\textwidth]{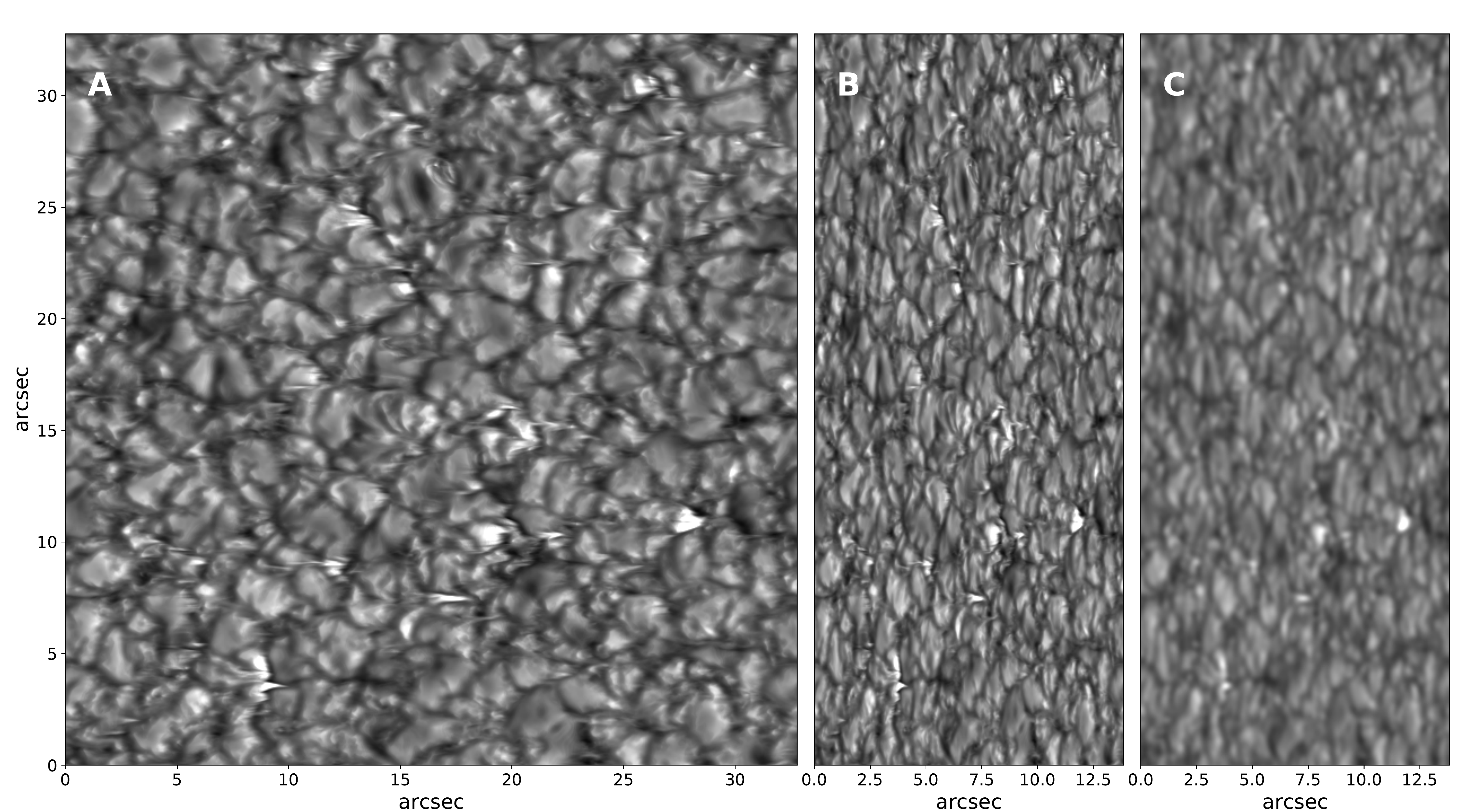}
\caption{Continuum intensity derived from the synthetic spectra at three different stages in the process. A: synthetic continuum image from the spectra emerging from a 65 $^{\circ}$ LOS at the native resolution of the simulation (i.e. 16 km sampling). B: continuum intensity after foreshortening the pixels along the LOS direction. C: continuum image after spatially convolving the synthetic data with an Airy disk of 240~km radius. \label{fig:continuum}}
\end{figure}

The next step in this process is the application of the spatial degradation. To first order, the Hinode/SP spatial PSF can be modeled by an circular Airy disk of 0.32". Although additional contributions due to the central obscuration, the spider and a defocus component have an impact on the intensity contrast of the SP data \citep{danilovic2008}, here we neglect these effects.
For each wavelength and Stokes parameter, we convolve the synthetic data with a 2D Airy disk with a $240$~km radius, reducing the continuum intensity contrast to 5.2\% (from 8.2\% in the foreshortened cube). The effects of pixel foreshortening and convolution with a spatial PSF can be seen in Fig. \ref{fig:continuum}.

The last step in the generation of simulated spectra is the addition of synthetic noise. The signal-to-noise ratio (SNR) of real Hinode/SP polar observations depends on the exposure time and the binning characteristics of the observations. Additionally, for any given polar map, the SNR changes as a function of the distance to the limb because the continuum intensity decreases quickly close to the edge of the solar disk. Analysis of a polar map taken on 2007 Sep 25 (at 00:10UT) reveals a noise level of $\sigma_I/I_C \sim 7\cdot 10^{-3}$ for the intensity and $\sigma_{QUV}/I_C= 8\cdot 10^{-4} – 1.2 \cdot 10^{-3}$ for the three states of polarization. Here, $\sigma_X$ is the standard deviation of the intensity or polarization signals in a range of continuum wavelengths, and $I_C$ is the average continuum intensity in that same range of wavelengths. For the purpose of this  investigation, we select a noise level of $\sigma_I/I_C= 7\cdot 10^{-3}$ for Stokes I, and $\sigma_{QUV}/I_C= 1\cdot 10^{-3}$ for Stokes Q, U, and V and we add synthetic gaussian noise of the equivalent amplitude to the spectra.

\subsection{Pixel binning}

The formulation of the magnetic filling factor is designed to quantify the fraction of the light that is of "non-magnetic" origin. This could be due to (unpolarized) stray light contamination at the detector, or to an unresolved non-magnetic component in the Sun's atmosphere within the geometric area covered by the pixel. Either way, when we do inversions with free filling factor, we are allowing the inversion code to tease out the magnetic component of the atmosphere and assign it a fraction of the pixel area. This simple modeling technique is designed to recover the intrinsic vector magnetic field from the observed spectra, instead of its apparent (i.e. pixel-averaged) counterpart.

In this paper, we compare the inverted magnetic field to that of the MURaM simulation in two ways. We first assess the apparent retrieved magnetic quantities (i.e. scaled by the magnetic filling factor) against the spatially degraded simulation (where the spatial PSF of Hinode/SP is also applied to the simulated quantities) to gauge how well the inversion codes do at quantifying the average magnetic properties. Secondly, we compare the intrinsic inverted magnetic field to the spatially resolved (undegraded) simulation to gain an understanding of how well the magnetic filling factor accounts for the effects of unresolved magnetic structure in the pixels.
For the second comparison it is necessary to retain the original spatial sampling of the synthetic spectra in order to compare the inversion results to the simulation at its native resolution. For this reason, we forego binning the synthetic data in the spatial domain, so the synthetic pixel size is much smaller than real Hinode/SP pixels, even though the spatial resolution is the same. 

\section{Spectral line inversion set up}\label{sec:inversion_setup}

We carry out the inversions of the $1536 \times 649 = 999,864$ sets of Stokes profiles of the Fe {\sc i} line pair with two Milne-Eddington inversion codes that model the non-magnetic atmospheric component in different ways. While one of the codes uses a global stray light profile to fit the non-magnetic part of the spectrum, the other one computes a local average intensity profile within a small radius (1") around each pixel to account for this non-magnetic component. There is some debate in the literature as to what approach is the best one to capture the intrinsic properties of the small-scale quiet Sun magnetism \citep{orozco2007PASJ, asensio_ramos}. Here we test both codes to determine whether one approach is more accurate than the other.

\subsection{MERLIN inversions}

The official Hinode/SP data pipeline at the High Altitude Observatory's Community Spectro-Polarimetric Analysis Center (CSAC) produces Milne-Eddington spectral line inversions of the entire Hinode/SP database. The Milne-Eddington gRid Linear Inversion Network \citep[MERLIN,][]{MERLIN} is an inversion code designed specifically to run automatically on Level 1 Hinode/SP calibrated spectra. It is written in C++ and the inversion scheme is based on the least-squares fitting of the observed Stokes profiles using the Levenberg-Marquardt algorithm. The inversion results are then packaged as a Level 2 data product and distributed via multiple outlets, including HAO's data download page\footnote{\url{https://csac.hao.ucar.edu/sp_data.php}} and Lockeed Martin Solar and Astrophysics Laboratory's website\footnote{\url{https://www.lmsal.com/hek/hcr?cmd=view-recent-events&instrument=sotsp}}.

The stand-alone source code for MERLIN can be found on the CSAC website (https://www2.hao.ucar.edu/csac/csac-spectral-line-inversions). Its main user-provided inputs are the observed Stokes spectra and a non-magnetic global stray light profile. Prior to the spectral line inversion, the CSAC data pipeline runs SolarSoft's \verb|SP_PREP| package on Level 0 SP data to produce calibrated Level 1 spectra as well as a number of quick-look data products. Among the latter is the scattered light spectral profile (\verb|SCATLP|), which is an average intensity profile over the regions of an SP map with a net polarization $P_{\rm tot} \leq 0.0035$ \citep{sp_prep}. This average {\em non-magnetic} intensity spectrum is unique for each SP map, and is used as the input scattered light profile for MERLIN in the Level 2 processing. 
In order to mimic the the CSAC data pipeline as closely as possible, we compute a global {\em non-magnetic} intensity profile from the synthetic SP map using only the pixels with low net polarization levels. Then, we provide this alongside the synthetic Stokes profiles as an input for the inversion.
The rest of the inversion parameters are kept the same as in the data pipeline (such as the inversion weights, the free parameters of the model, and the hard limits on the inverted atmospheric quantities).

\subsection{MILOS inversions}
The MILne–Eddington inversion of the pOlarized Spectra \citep[MILOS,][]{MILOS} code is used by the Nagoya University group for the generation of vector magnetic field maps of the polar regions observed by Hinode/SP\footnote{doi: 10.34515/DATA.HSC-00001}. This database, published in 2022, is a rather complete collection of $\sim 400$ inverted and disambiguated Hinode polar maps going back to 2012.

Instead of a global scattered light profile, MILOS uses a locally-computed stray light spectrum to represent the {\em non-magnetic} component of each pixel in the observation. \cite{orozco2007PASJ} deemed the local scattered light profile with variable filling factor necessary to account for the effects of telescope diffraction in order to accurately retrieve quiet sun magnetic fields. In polar observations, the rapid decrease of intensity as one approaches the limb (i.e. limb-darkening) lends further justification to this local stray light approach. It is worth noting here that the synthetic data generated for this work do not present limb-darkening, as the emergent spectrum for every pixel in the FOV was calculated for the same viewing angle of 65${^\circ}$. Following \cite{orozco2007PASJ, tsuneta2008, Ito2010}, the stray-light profile is evaluated for each pixel individually and computed as the average Stokes I profile in a 1''-wide box centered on the pixel under consideration. We implement an analogous procedure on the MURaM spectra and generate a local stray light matrix that is given as an input to MILOS along with the synthetic observations.

\section{Results}\label{sec:inversion_results}

In this section, we compare the inversion results (in particular the vector magnetic field) delivered by the two codes against each other and against the ground truth of the MURaM simulation. We study separately the intrinsic and the apparent ("pixel-averaged") magnetic quantities.

\subsection{MERLIN vs. MILOS inversion results}

\begin{figure}[t!]
\centering
\includegraphics[width=1.0\textwidth]{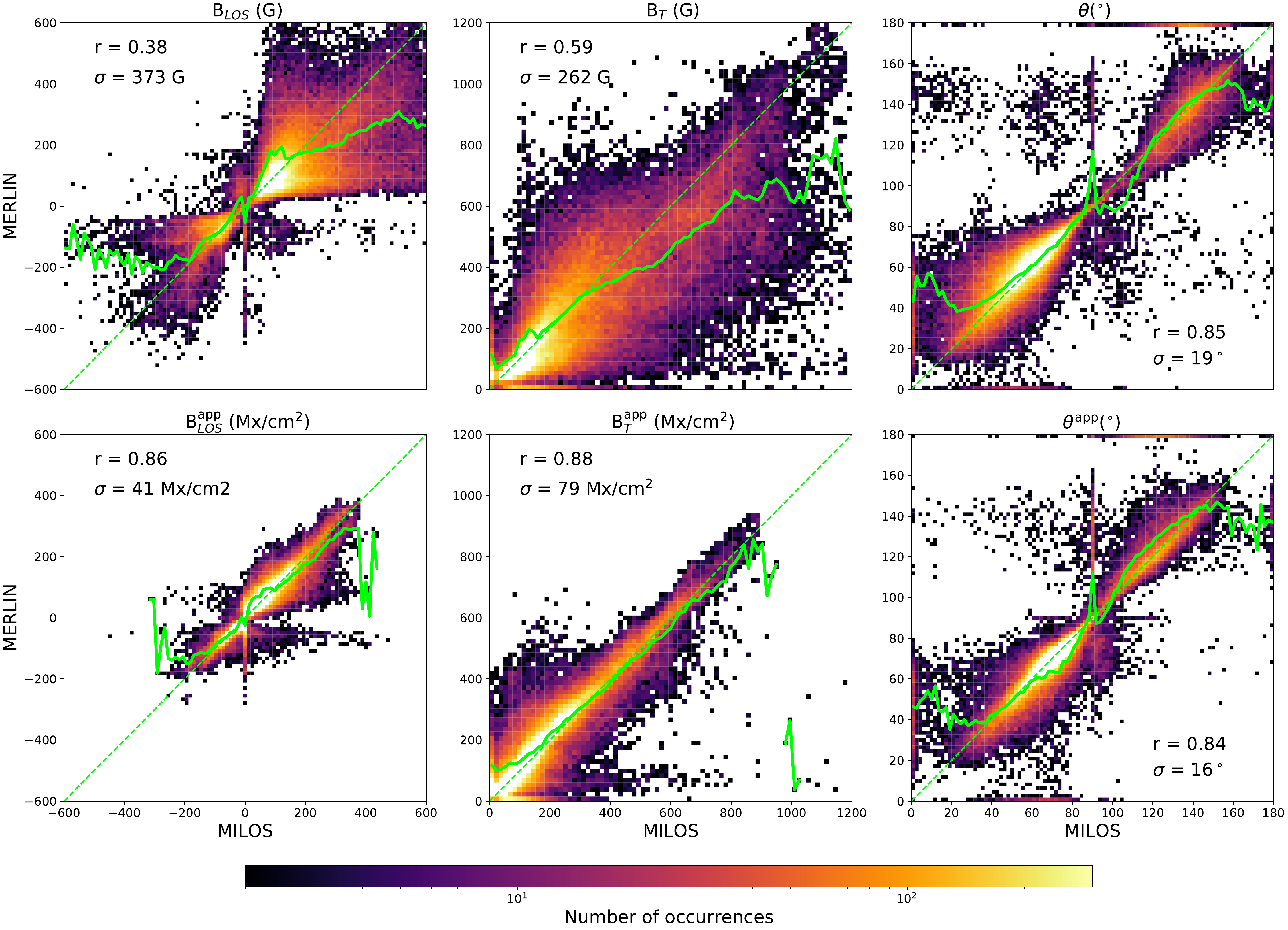}
\caption{Comparison of retrieved magnetic quantities from MILOS and MERLIN inversions. Each panel contains a density scatter plot of MERLIN against MILOS results for a specific magnetic quantity. The top row corresponds to the intrinsic magnetic field components, while the bottom row shows their corresponding apparent counterparts. From left to right: LOS component of the magnetic field, transverse component of the magnetic field, magnetic field inclination with respect to the LOS. Note that the colorbar represents the number of occurrences on a logarithmic scale. The solid green line shows the average MERLIN value in each bin on the MILOS x-axis. The dashed-line marks the ideal one-to-one relationship. \label{fig:milos_vs_merlin}}
\end{figure}

First, we compare the inversion results delivered by the two codes. 
Because the synthetic spectra are noisy, it makes sense to only look at those pixels with total polarization levels above a certain threshold in order to avoid spurious results due to the inversion interpreting the noise. We use the total polarization, P$_{\rm TOT}$:

\begin{equation}
P_{\rm TOT}= \sum_{\lambda} \sqrt{Q^2+U^2+V^2}/(I_{\rm C} N_{\lambda})
\end{equation}

as the metric to select pixels. Here, $N_{\lambda}$ is the number of wavelengths in the integrating window. 
The benchmark noise level is the total polarization computed in a wavelength window in the continuum ($P_{\rm CONT}$). In order to consider a pixel for further analysis, the total polarization within the spectral lines must exceed a polarization threshold that is five times above the noise level ( $P_{\rm TOT} \geq 5 P_{\rm CONT}$).

Fig \ref{fig:milos_vs_merlin} presents density scatter plots of the inverted magnetic quantities, with MILOS on the x-axis and MERLIN on the y-axis. Only pixels with total polarization above the threshold are considered. From left to right, the columns correspond the LOS magnetic field, the transverse magnetic component and the inclination angle with respect to the LOS. The bin sizes are 12~G for the field components and 1.8$^{\circ}$ for the inclination. The top row looks at the intrinsic magnetic quantities whilst the bottom row highlights their apparent counterparts. The solid green lines show the average value retrieved by MERLIN against the MILOS results in each bin.

The Pearson correlation between the MERLIN and MILOS results, $r$, as well as the standard deviation of their differences, $\sigma$, are stated in each panel. On average, the apparent magnetic quantities agree much better than the intrinsic ones. The scatter plots of the apparent magnetic fluxes (bottom row) show higher correlations (larger $r$ values) and are better aligned with the one-to-one correspondence (dashed lines) than the intrinsic quantities in the top row. The distributions of the former also exhibit more compactness owing to the lower $\sigma$ values.
The azimuth angle from the two inversions (not shown in the figure) agrees very well, with a Pearson correlation of $r=0.89$ for the quantity $sin(\chi)$ and a standard deviation of $\sigma=8.6^{\circ}$. There are no visible systematic offsets between the azimuths retrieved by the two codes.

\begin{figure}[t!]
\centering
\includegraphics[width=0.49\textwidth]{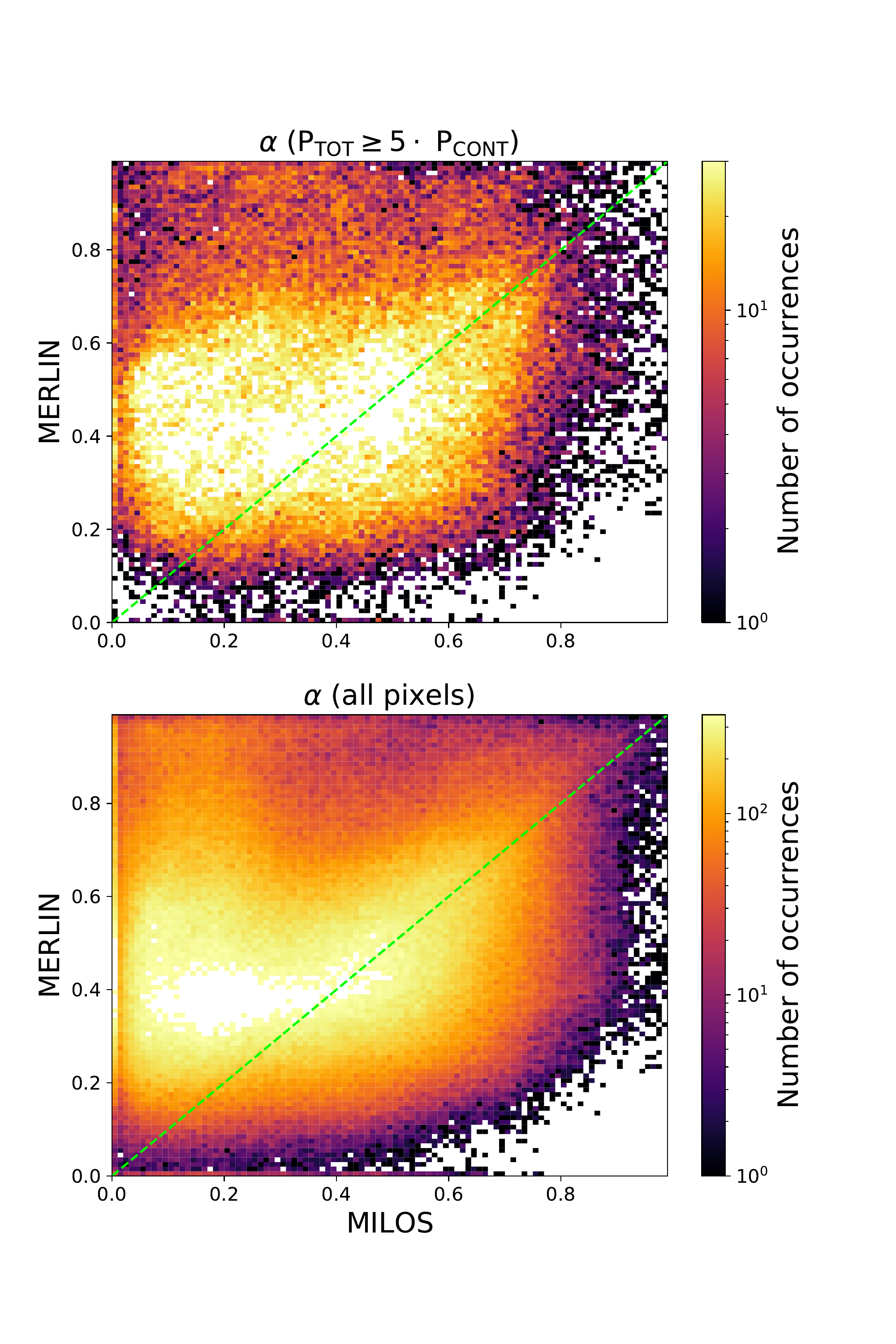}
\caption{Density scatter plot of the magnetic filling factors retrieved by MERLIN and MILOS. On the top, we show only the pixels above the polarization threshold. The bottom panel shows the scatter plot for all pixels in the FoV.\label{fig:milos_vs_merlin_alpha}}
\end{figure}

The histograms of the top row show MILOS' systematic preference for stronger intrinsic fields (both in the LOS and the transverse components) when compared with MERLIN. 
The relative biases in the intrinsic field strengths derived by MILOS and MERLIN can be attributed to the systematic differences in the retrieved magnetic filling factor (upper panel of Fig. \ref{fig:milos_vs_merlin_alpha}). While MILOS has a preference for lower values of $\alpha$, with an average around $\sim 0.3$, MERLIN tends to retrieve a more even distribution with an average around $\alpha \sim 0.5$. This difference between the two inversion results holds true for all pixels, not only those exceeding the polarization threshold (see lower panel of the same figure). It is mainly due to the differences in the treatment of the stray light between the two codes: while MILOS considers a local stray light profile that is different for each pixel in the FOV, MERLIN, on the other hand, uses a globally-computed stray light (the same for all pixels) and allows it to freely shift in wavelength to better fit the observed spectra (i.e. "the stray light shift" is an additional free parameter of the inversion model). 
Although delving into this is out of the scope of this paper, our preliminary tests point to the "stray light shift" as the main cause of the disagreement between the two inversions, rather than the local vs. global nature of the stray light profile itself.

\subsection{Apparent magnetic quantities vs. spatially-degraded MURaM}\label{apparent_vs_muram}

In this section we compare the apparent magnetic parameters retrieved by the inversions with the simulated quantities from MURaM. 

The concept of {\em magnetic filling factor} does not have meaning in the context of the MURaM cube. At each spatial location of the simulation grid there is one set of physical properties with a single value of the magnetic field vector. 
In the interpretation of real observations the magnetic filling factor is typically used to explain a combination of instrumental stray light and spatial coupling due to the finite angular resolution of the telescope. Because the synthetic spectra in this work do not contain stray light contamination, the inversion model with variable magnetic filling factor can only be used to explain the mixing of the signal between the adjacent pixels due to effects of the spatial degradation with a 0.32" PSF. In other words, the model will attempt to capture the intrinsic values of the unresolved magnetic fields in the resolution element of the synthetic observation.

Due to the non-linearity of the radiative transfer process, the average spectrum emerging from a region of the Sun is not identical to the spectrum calculated from the average atmospheric properties in that same region. Therefore we must be careful when interpreting the results of the inversions of spatially averaged signals as the average atmospheric properties. Nevertheless, in this section we carry out a comparison between the apparent magnetic quantities inferred from the inversion with the spatially-degraded simulation quantities. We will show that the former meaningfully represent the latter when the simulation is degraded to the spatial resolution of the synthetic observations. This approach was defended by \cite{plowman+berger2, plowman+berger3} for comparisons between simulated GONG magnetic flux data to its respective MURaM ground truth.

In order to carry out a quantitative one-to-one comparison between the inversion results and the ground-truth MURaM atmospheres, we first foreshorten the pixels  
in the sheared simulation box along the same direction and by the same factor we did for the spectra. Then, we apply the same spatial degradation (i.e. a 2D convolution with an 0.32" Airy disk) at each height in the foreshortened MURaM cube. These operations are applied onto the three cartesian components of the vector magnetic field separately and the result is then transformed to spherical coordinates for comparison with the inversion results. Lastly, the MURaM azimuths are folded to range between $0 - 180^{\circ}$ in order to mimic the 180$^{\circ}$ ambiguity in the azimuths retrieved by the inversion codes.

\begin{figure}[t!]
\centering
\includegraphics[width=1.0\textwidth]{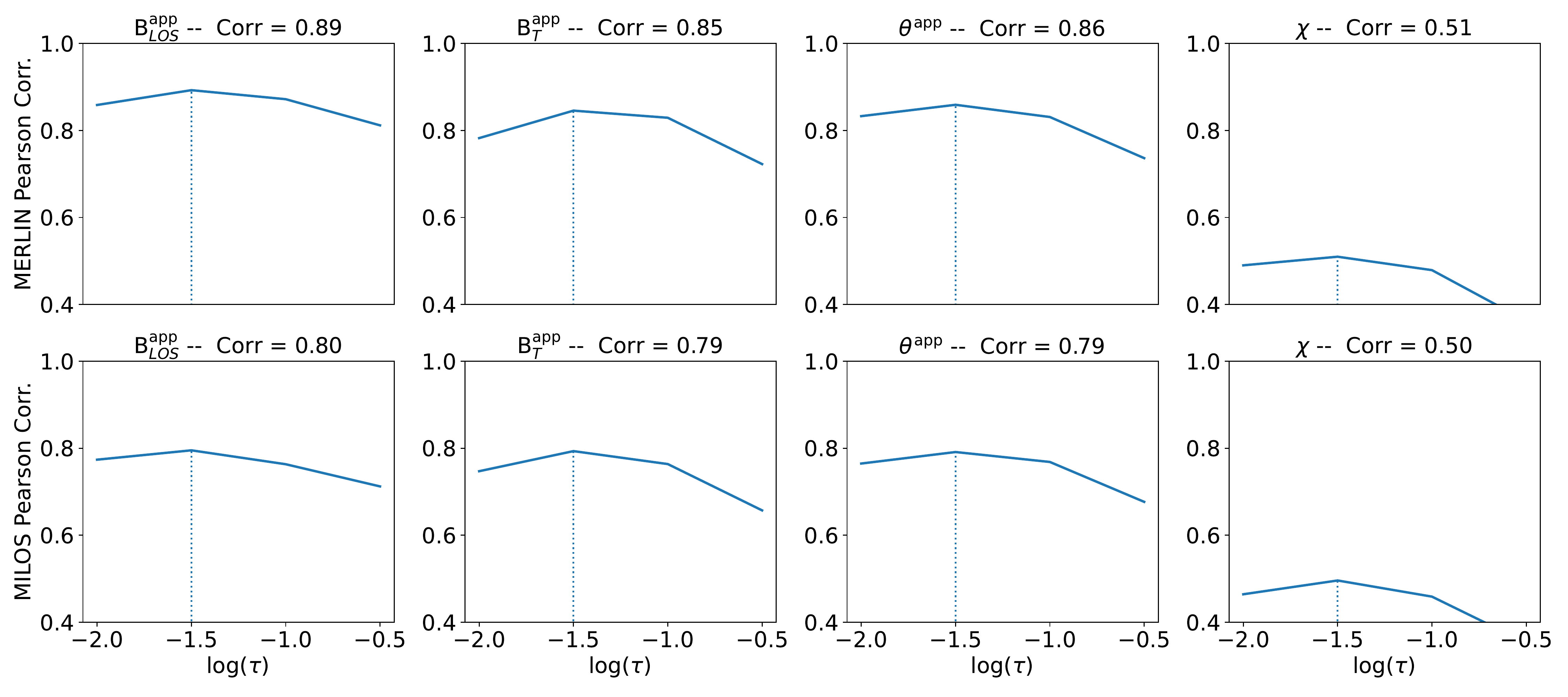}
\caption{Pearson correlation between the apparent vector magnetic field parameters delivered by the inversions and the degraded MURaM simulation, as a function of optical depth in the simulation ($log(\tau) = -0.5, -1, -1.5, -2$). The top row shows the MERLIN results whilst the bottom row concerns itself with MILOS inversions. \label{fig:correlations_apparent_vs_muram}}
\end{figure}

For each synthetic Stokes vector, Milne-Eddington inversions produce a single realization of the vector magnetic field, which is representative of the atmospheric properties around the height of maximum sensitivity of the spectral lines to this physical parameter \citep{westendorp_plaza}. To find the average optical depth probed by the inversions, we study the correlation between the inverted quantities and the simulated properties at different horizontal slices (at log$(\tau)$= -0.5, -1, -1.5, -2) in the spatially-degraded MURaM cube. Fig. \ref{fig:correlations_apparent_vs_muram} shows the Pearson correlation coefficient between the {\em apparent} inverted magnetic parameters and their spatially-degraded MURaM counterparts as a function of optical depth in the simulation. The top row corresponds to the MERLIN results and the bottom row to the MILOS inversions. The value of the maximum correlation in each case is stated in the panel title. Overall, the MERLIN results show a better agreement with the simulation than the MILOS ones. However, the maximum sensitivity of the Milne-Eddington inversions to the magnetic field happens at log$(\tau)\sim -1.5$ along the line of sight, regardless of the inversion code. 
The magnetic field azimuth (last column) shows the worst correlation. As will become clear from the scatter plots of the next figure, this is mostly an effect of the $180^{\circ}$ ambiguity.

From this point onwards, we take log$(\tau)\sim -1.5$ as representative of the optical depth 
probed by the inversions at this $\mu$ (in what relates to the magnetic field). It is worth noting that the sensitivity to other physical parameters, such as the LOS velocity, may reach a maximum at different heights.

\begin{figure}[t!]
\centering
\includegraphics[width=1.0\textwidth]{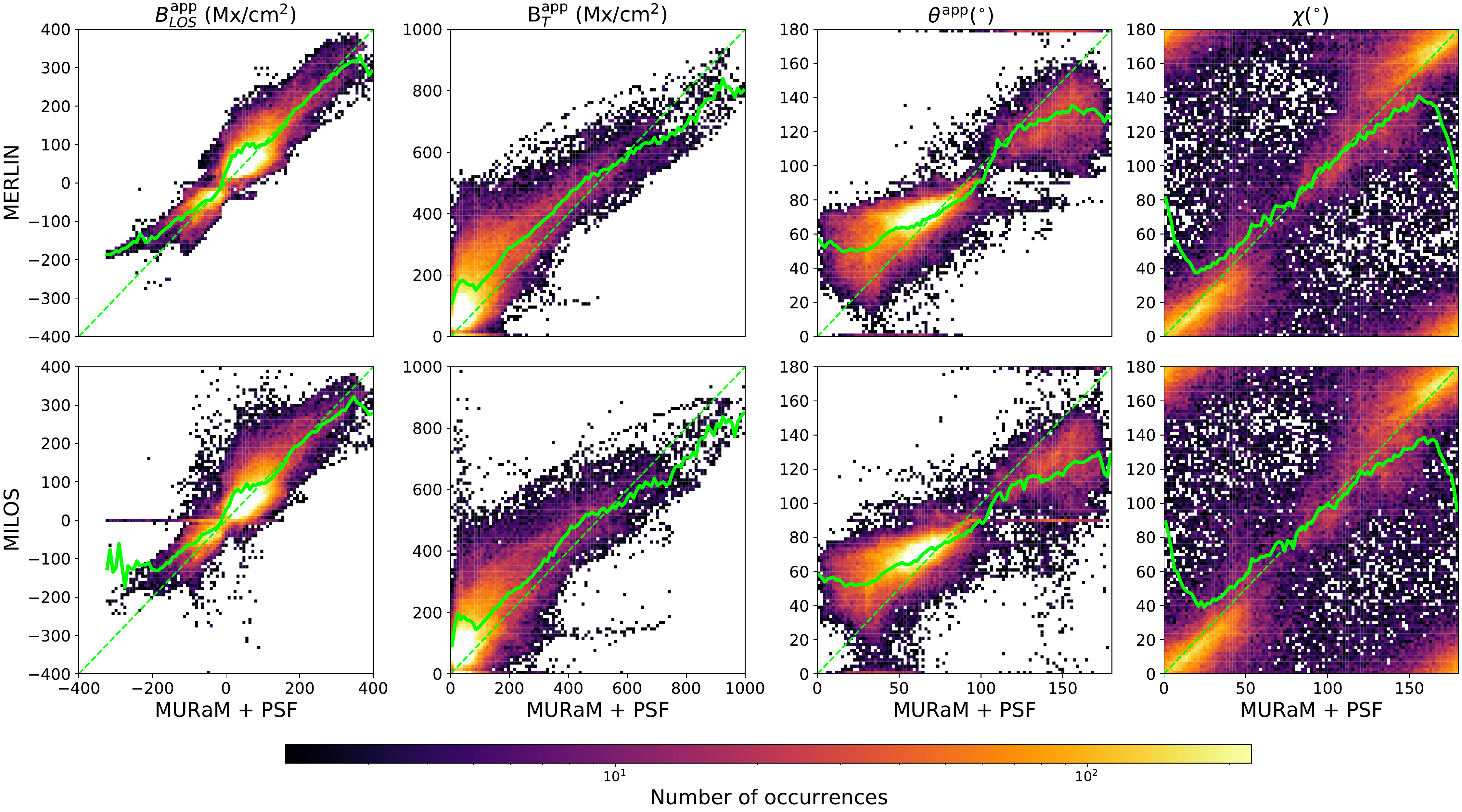} 
\includegraphics[width=1.0\textwidth]{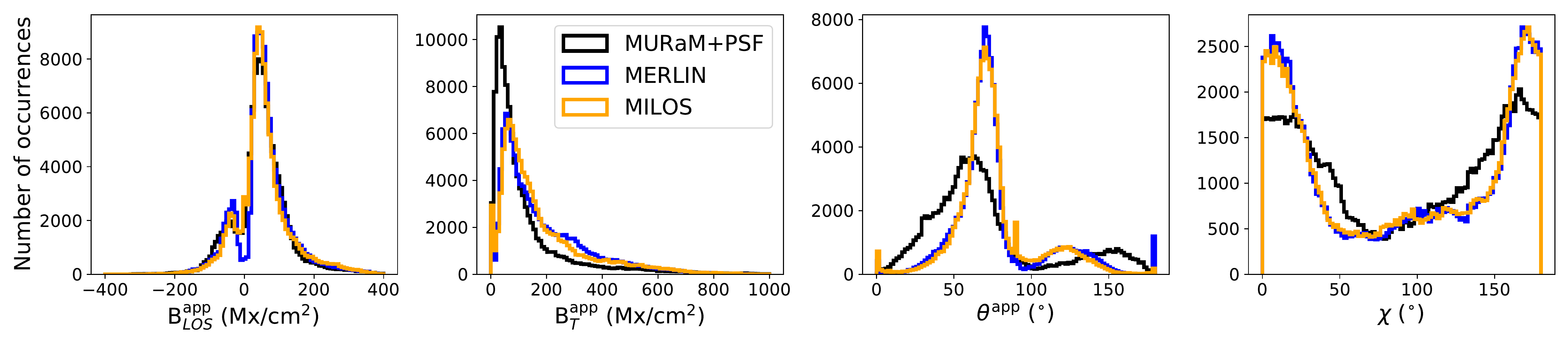}
\caption{Comparison between the {\em apparent} magnetic quantities retrieved by the inversion and the magnetic field components in the spatially-degraded MURaM cube at log$(\tau=-1.5)$. The top (middle) row shows the density scatter plots of the MERLIN (MILOS) inverted quantities against their MURaM counterparts. The bin sizes are 8~Mx/cm$^2$, 10~Mx/cm$^2$, and 1.8$^{\circ}$ for the LOS and transverse fluxes, and the angles, respectively. The solid and dashed green lines have the same meaning as in Fig. \ref{fig:milos_vs_merlin}. The bottom row of the figure shows 1D histograms of the same {\em apparent} magnetic components, with the spatially-degraded MURaM in black, MERLIN in blue and MILOS in orange.  \label{fig:apparent_vs_muram}}
\end{figure}

The first two rows of Fig. \ref{fig:apparent_vs_muram} show density scatter plots of the {\em apparent} magnetic components against their counterparts in the spatially-degraded MURaM simulation at log$(\tau)\sim -1.5$. The top and middle rows correspond to MERLIN and MILOS inversions, respectively. From left to right, the inverted quantities (y-axes) correspond to the {\em apparent} LOS and transverse magnetic fluxes, the apparent field inclination and the azimuth in the plane of the sky. The size of the bins are 8~Mx/cm$^{2}$, 10~Mx/cm$^{2}$ for the fluxes, and $1.8^{\circ}$ for the angles.
The {\em apparent} LOS flux density (first column) retrieved by both inversions agrees well with the spatially degraded LOS field component of MURaM, although for very low field strengths, the inversions tend to overestimate it. The second column of Fig. \ref{fig:apparent_vs_muram} portrays a general over-estimate of the {\em apparent} transverse flux density, which is more pronounced in the weak field regime. This is most likely an inversion bias due to the (synthetic) photon noise in the linear polarization profiles. The {\em apparent} magnetic field inclination (third column) retrieved by the inversions also suffers from a systematic bias towards more inclined fields (i.e. $\theta \sim 90^{\circ}$) than the simulation ones.
Lastly, the inverted azimuth (fourth column) generally agrees with the simulated values, albeit with a generous spread around the ideal one-to-one relation (dashed green line). The uncertainty in the derived azimuth is typically a direct consequence of noise in Stokes $Q$ and $U$. The top left and bottom right corners of the azimuth panels show excess density due to the $180^{\circ}$ azimuth ambiguity. This skews the average inversion value (solid green line) away from the one-to-one relation (dashed line) and misleads the Pearson correlation coefficients returned by Fig. \ref{fig:correlations_apparent_vs_muram}.

The bottom row of Fig. \ref{fig:apparent_vs_muram} complements the scatter plot information with 1D histograms. The distributions of {\em apparent} magnetic quantities derived by MERLIN (blue) and MILOS (orange) are overplotted on the distributions of the magnetic components in the spatially-degraded MURaM cube (black). While $B_{\rm LOS}^{\rm app}$ represents well the LOS magnetic field component in the degraded simulation, the inversions tend to overestimate the amount of strong {\em apparent} transverse flux density and skew the apparent inclination angle towards $90^{\circ}$. The retrieved azimuths deviate from a uniform distribution, showing an excess population around $0^{\circ}$ and $180^{\circ}$. Albeit less pronounced, this feature is also present in the spatially-degraded MURaM simulation. This corresponds to excess of transverse fields oriented along the foreshortening direction, and it is mainly due to the projection effect of a highly inclined LOS on mostly radial magnetic fields. We explain the origin of this bias in the Appendix.

\subsection{Intrinsic magnetic quantities vs. undegraded MURaM}\label{intrinsic_vs_muram}
\begin{figure}[t!]
\centering
\includegraphics[width=1.0\textwidth]{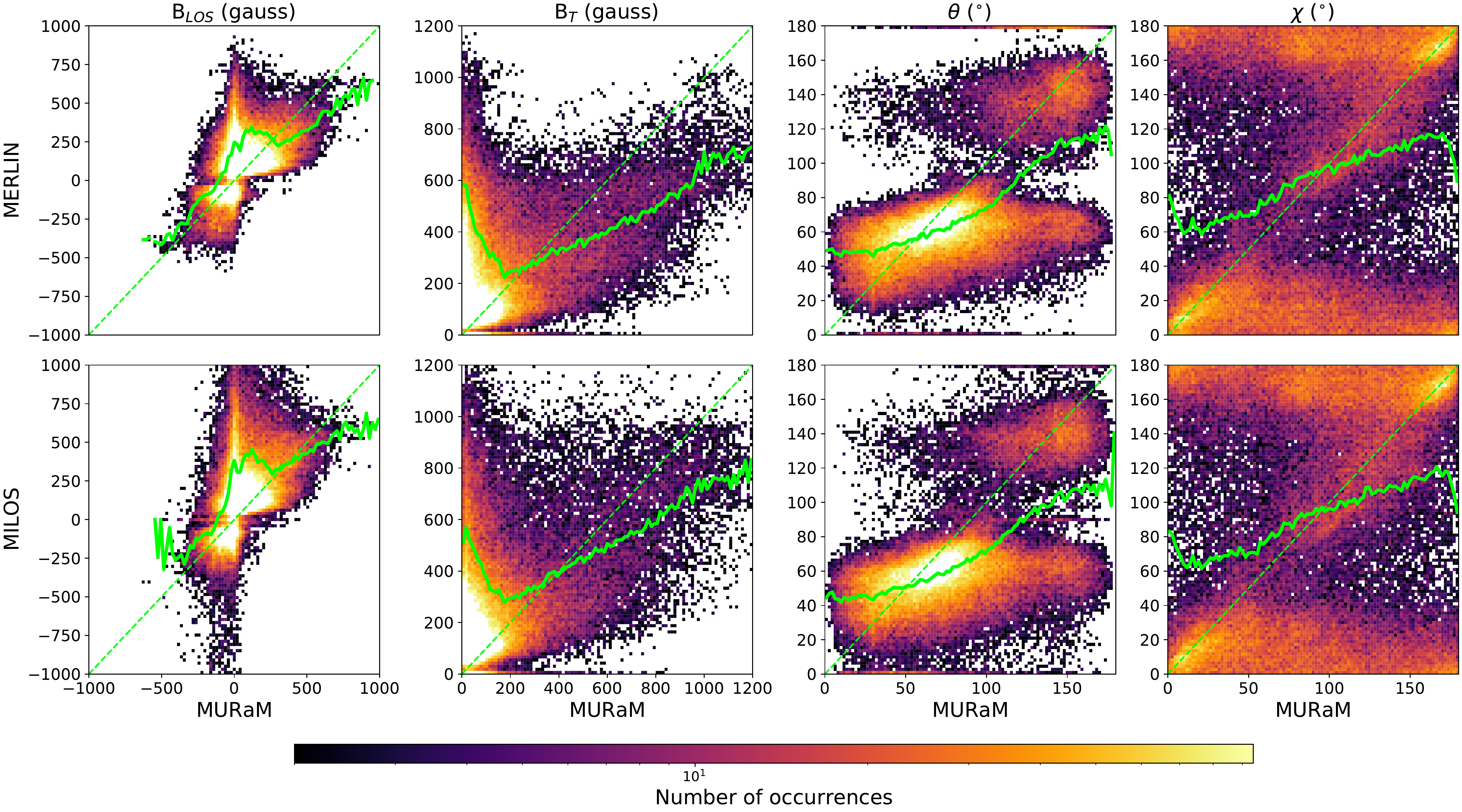} 
\includegraphics[width=1.0\textwidth]{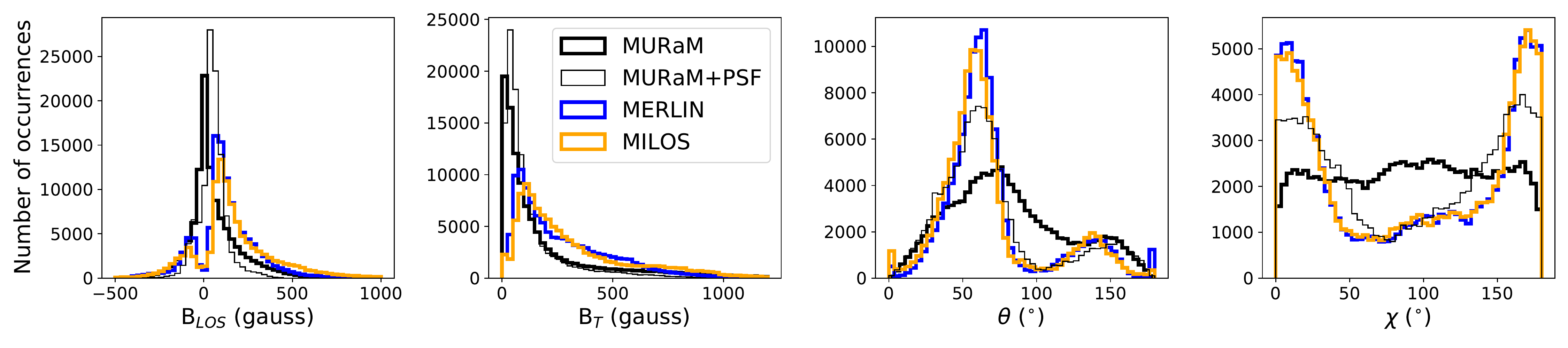}
\caption{Comparison between the {\em intrinsic} magnetic quantities retrieved by the inversions and the magnetic field components in the foreshortened (but not spatially-degraded) MURaM cube. The top (middle) row shows the density scatter plots of the MERLIN (MILOS) inverted field against its MURaM counterpart. The bin sizes are 20~G, 12~G, and 1.8$^{\circ}$ for the LOS and transverse fields, and the angles, respectively. The solid and dashed green lines have the same meaning as in Fig. \ref{fig:milos_vs_merlin}. The bottom row of the figure shows 1D histograms of the same magnetic field components with MURaM in thick black lines, MERLIN in blue and MILOS in orange. The histograms of the spatially degraded MURaM simulation are also plotted for comparison (thin black line).  \label{fig:intrinsic_vs_muram}}
\end{figure}

In this section we compare the intrinsic magnetic parameters retrieved by the inversion to the magnetic field in the foreshortened MURaM cube. In this case, no spatial degradation is applied to the simulation box.
With this approach we evaluate the usefulness of a magnetic filling factor in the atmospheric model underlying the spectral line inversion. In particular, we assess if $\alpha$ is able to model the effect of the point spread function (i.e. the mixing of spatial information) on the emergent Stokes spectra, and whether or not the retrieved {\em intrinsic} magnetic field values are a good representation of the physical conditions in each MURaM pixel at its native foreshortened resolution. 
Fig. \ref{fig:intrinsic_vs_muram} is analogous to Fig. \ref{fig:apparent_vs_muram}, but now, instead of comparing the {\em apparent} inverted magnetic field to the spatially-degraded simulation, we compare the {\em intrinsic} magnetic properties to the un-degraded (yet foreshortened) MURaM cube. Again, we take only the pixels with $P_{\rm TOT} > 5 P_{\rm CONT}$. A correlation study analogous to that of Fig. \ref{fig:correlations_apparent_vs_muram} revealed very weak Pearson correlations between the inversion results and the MURaM simulation for all optical depths (with typical values lying around $r \sim 0.5$). For this reason, we chose to compare the inversion results to the same horizontal slab of the simulation ($log(\tau) = -1.5$) as in the previous section.

The scatter plots of Fig. \ref{fig:intrinsic_vs_muram} show a bleak picture of what the inversion codes are able to accomplish. The retrieved magnetic properties only marginally resemble the ground truth of the simulation. In particular, the LOS and transverse field components are generally overestimated by the inversion, especially in the weaker field regime, while the inclination and azimuth angles present weak correlation to the MURaM values.

The 1D histograms in the third row provide an alternative view of the same data. While the MURaM distributions of the LOS and transverse field strengths (solid black lines in the first two panels) peak at 0~G, both inversion codes completely miss these very weak fields. Additionally, the inversions seem to be blind to the population of purely transverse fields (i.e. the missing $\theta \sim 90^{\circ}$ population in the third panel), and the azimuth retrieval is heavily skewed towards the edges of the distribution ($0^{\circ}$ and $180^{\circ}$, last panel) compared to their undegraded MURaM counterparts (thick black lines). These biases exist in the spatially-degraded simulation too (compare the thin and thick black lines in the 1D histograms):
the effect of convolving the foreshortened MURaM cube with Hinode's PSF results in a loss of purely transverse fields ($\theta = 90^{\circ}$) and skews the azimuth values towards $0^{\circ}$ and $180^{\circ}$ in the simulated quantities. The inversion results are further impacted by the presence of synthetic noise as well as the limitations of the Milne-Eddington modeling assumptions, making some of these effects even more pronounced.

It is interesting to note that the  azimuth distribution of the spatially-degraded MURaM simulation (thin black line in Fig. \ref{fig:intrinsic_vs_muram}) shows the same biases as the inversion results, whilst the undegraded simulation presents an almost uniform distribution of azimuths. This is merely due to a selection bias imposed by the polarization threshold. One of the effects of convolution with a spatial PSF is the "dilution" and enlargement of the strongly magnetized patches. Because the polarization threshold is defined on the spatially-degraded spectra, it results in a thresholding mask that lets weakly magnetized pixels of the undegraded simulation make it into the PDF. These weakly magnetized pixels have more randomly oriented magnetic fields that result in a uniform distribution of azimuths, explaining the thick black line in the last panel of Fig. \ref{fig:intrinsic_vs_muram}. See the Appendix for an in-depth analysis of this effect.

\section{Conclusions}


The measurement from Earth of the radial flux at the solar poles is severely curtailed by foreshortening and projection effects, as well as by the differential sensitivity of the linear and the circular polarizations to the magnetic field \citep{petrieLRSP}. 
Additionally, the inference of the vector magnetic field through the Zeeman effect is subject to a 180$^{\circ}$ azimuth ambiguity in the plane of the sky (i.e. there are two solutions for the transverse component of the magnetic field that are separated by 180$^{\circ}$). This, in turn, leads to two solutions for the inclination of the magnetic field in the local reference frame, and additional physical assumptions need to be used in order to choose one solution over another \citep{metcalf2006, leka2009}. 
In order to compute these two solutions in the first place, one has to start from the {\em intrinsic} magnetic components estimated through the spectral line inversion process. The transformation relies on the inclination and azimuth angles in the local reference frame as well as the heliocentric angle of the observation \citep[see equations 1 and 2 in][]{Ito2010}.

In this work we carry out Milne-Eddington inversions of synthetic Stokes spectra calculated from a MURaM simulation of a plage region emulating a polar LOS of $65^{\circ}$. The synthetic spectra are degraded according to the specifications of Hinode/SP: with LOS foreshortening, a spatial resolution of 0.32", spectral sampling of 21.5 m\AA\ and noise levels typical of polar Hinode observations. The spectra are inverted with the MERLIN and MILOS inversion codes, and the retrieved magnetic fields are compared to those in the simulation cube.

While the {\em apparent} (i.e. "pixel-averaged") inverted magnetic quantities highly correlate with those of the {\em spatially degraded} simulation, the {\em intrinsic} magnetic properties present strong biases in the retrieved inclination and azimuth when compared to their counterparts in the {\em undegraded} MURaM cube.
The comparison of the retrieved intrinsic quantities to the undegraded MURaM simulation is eye-opening. The histograms of Fig. \ref{fig:intrinsic_vs_muram} suggest that the MILOS and MERLIN inversions tend to:
\begin{itemize}
    \item underestimate the number of pixels with no LOS field component,
    \item retrieve stronger transverse fields than are present in the simulation,
    \item severely underestimate the number of purely transverse ($\theta=90^{\circ}$) fields, and
    \item overestimate the number of pixels with transverse component parallel to the foreshortening direction ($\chi = 0^{\circ}$ or $\chi = 180^{\circ}$).
\end{itemize}

\noindent Biases in the {\em intrinsic} magnetic quantities will propagate to the calculation of the magnetic field inclination in the local reference frame, thus impacting the inference of the photospheric radial flux at the poles.

Photon noise is responsible for some of the systematic biases in the magnetic field retrieval, in particular for those quantities that depend on the linear polarization \citep{juanma2011, juanma2012}. Projection and foreshortening effects due to observing the solar poles from the Earth's perspective also play a role in some of the systematic inference errors. At a $65^{\circ}$ viewing angle, foreshortening accounts for a factor of $\sim 2$ spatial degradation along the foreshortening direction (compared to disk center observations). This effect will be significantly aggravated the closer the observations are to the limb. 
Convolution with the telescope's PSF seems to have a large effect on the inference biases ---
the distributions of the magnetic field vector in the spatially degraded MURaM cube (thin black lines in the bottom row of Fig. \ref{fig:intrinsic_vs_muram}) show the same biases (albeit less pronounced) as the inversion inferences, when compared against the distributions of the magnetic properties in the undegraded simulation (thick black lines in the same panels).
Although the inversion inferences from spatially degraded spectra are not expected to be equivalent to the inferences from spectra emerging from the spatially-degraded simulation, the apparent inversion results show very much the same trends as the distributions of the spatially-degraded MURaM. The reader is referred to the Appendix for a detailed analysis of the effects of LOS projection, pixel foreshortening and spatial degradation on the magnetic properties of the simulation.

{\em Our results show that a Milne-Eddington atmosphere with variable magnetic filling factor is not an adequate modeling technique for characterizing the spatial mixing that arises as a combined effect of LOS foreshortening and the telescope PSF, at least in the case where the magnetic fields exist at much smaller spatial scales than the size of the PSF, such as is the case of the MURaM simulation used in this work.}

Existing literature had already pointed in this direction. Modeling relatively-weak unresolved magnetic structures using a variable filling factor can result in systematically erroneous inferences \citep{pevtsov2021}. 
Additionally, when the non-magnetic component is modeled as a local stray light profile computed from the observations, correlations between the former and the latter can result in artificially small magnetic filling factors \citep{asensio_ramos}, and therefore artificially large intrinsic fields. This could be contributing to the systematic differences in the intrinsic magnetic fields obtained from MILOS and MERLIN.
Another consideration is the inherent limitations of the Milne-Eddington approximation itself.
Because the Milne-Eddington model is not able to generate asymmetries in the emergent Stokes profiles \citep{josecarlos_book}, it is possible that the inversion algorithm uses the stray light profile to fit asymmetries in Stokes I. More generally, the synthetic spectra used in this work are calculated from atmospheres with significant magnetic field and velocity gradients that cannot be captured by a simple Milne-Eddington model. 

More sophisticated inversions (such as SIR) that are naturally able to model spectral line asymmetries due to gradients along the LOS might do a better job at retrieving the intrinsic magnetic properties hidden within the resolution element of the observations.
 The most straightforward avenue for improving photospheric inferences is to increase the spatial resolution of the observations until we fully resolve the magnetic structures. But because the foreshortening effects at the poles can be large, there's an inherent limitation to what is achievable when observing from the ecliptic viewpoint.
 It is possible that observations of the polar chromospheric magnetic field (exploiting the power of Hanle-effect diagnostics) help bypass some of the biases produced by the finite spatial resolution, as the magnetic properties in the chromosphere are expected to be more uniform than those of the photosphere below it. But of course, when thinking of the solar poles, nothing beats observations from outside of the ecliptic \citep[such as those from the Solar Orbiter,][]{SOLO} and most ideally from a direct polar viewpoint.

\begin{acknowledgements}
This material is based upon work supported by the National
Center for Atmospheric Research, which is a major
facility sponsored by the National Science Foundation under
Cooperative Agreement No. 1852977. RC and NN acknowledge support from NASA LWS Award 80NSSC20K0217. IM and XS acknowledge support from NASA HGIO award 80NSSC21K0736. We would like to acknowledge high-performance computing support from Cheyenne (doi:10.5065/D6RX99HX) provided by NCAR's Computational and Information Systems Laboratory, sponsored by the National Science Foundation. In this work we have made use of Python packages NumPy \citep{numpy}, matplotlib \citep{matplotlib}, SciPy \citep{scipy} and Astropy \citep{astropy}.

\end{acknowledgements}

\appendix

The projection of mostly radial fields onto a LOS reference frame results in vectors with transverse components that are predominantly oriented towards (or away from) the observer. This is the reason for the strong skew in the distributions of magnetic field azimuth in Figures \ref{fig:apparent_vs_muram} and \ref{fig:intrinsic_vs_muram}. Here, we show step by step the effects of a slanted viewing angle, pixel foreshortening and spatial degradation in the distribution of magnetic field azimuths for the idealized scenario of a circular magnetic patch with azimuthally symmetric magnetic fields ("azimuth center").

\begin{figure}[t!]
\centering
\includegraphics[width=1.0\textwidth]{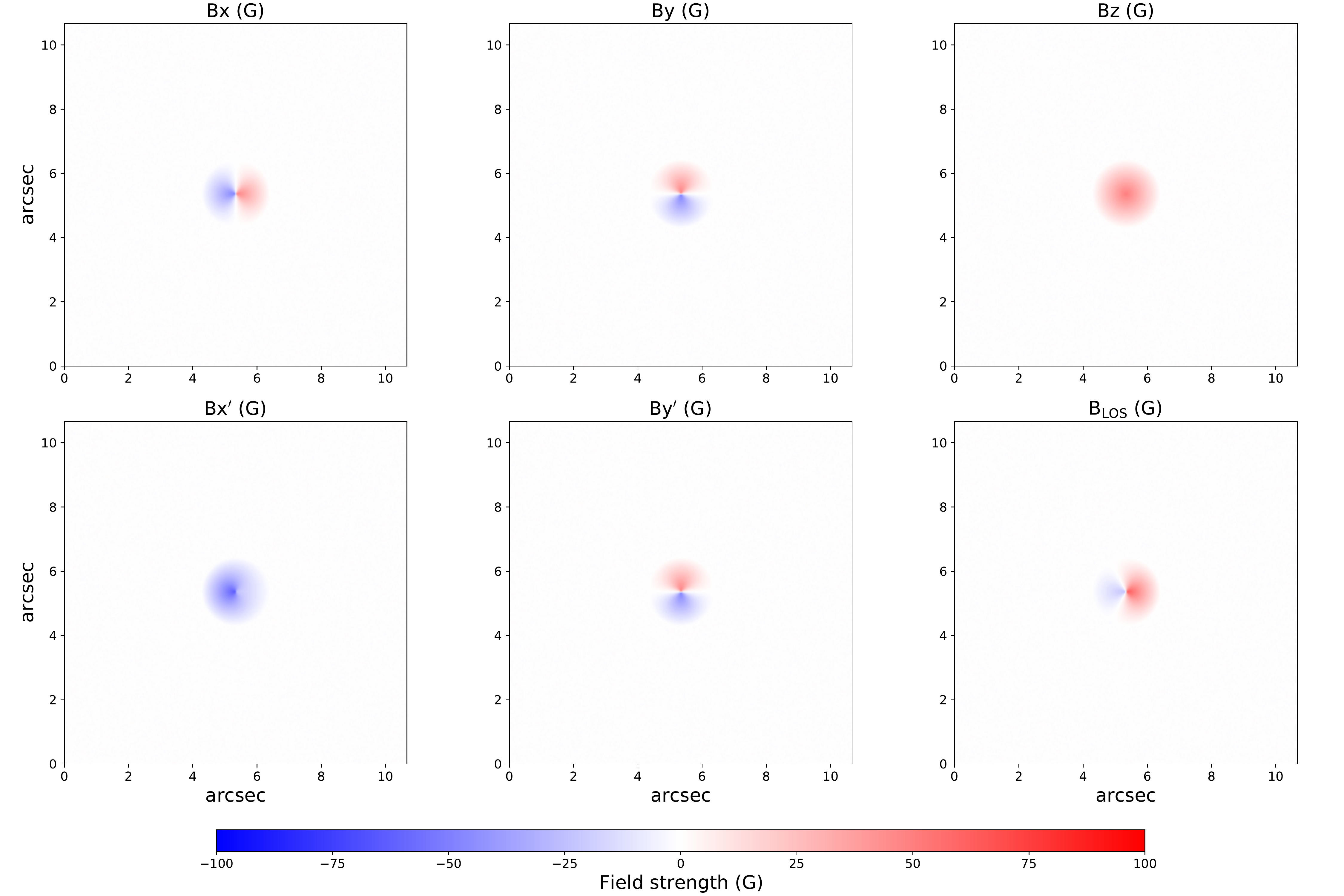}
\caption{Magnetic field components of the magnetic patch. Top: in the original ("local") reference frame. Bottom: in the LOS reference frame.  \label{fig:azim_experim_B}}
\end{figure}

First, we generate a 2D circular magnetic patch with a field strength that decreases radially inside the patch, and whose horizontal component points azimuthally outwards from the center (top row of Fig. \ref{fig:azim_experim_B}). Here, $B_z$ represents the local vertical component of the magnetic field vector, and $B_x$ and $B_y$ its components tangential to the solar surface. The magnetic field outside of the circle follows a normal random distribution with very weak strengths.
Next, we simulate observing this magnetic patch from the left side, with an heliocentric viewing angle of $65^{\circ}$. We transform the vector magnetic field at each pixel onto the LOS reference frame to obtain $B_{\rm LOS}$ and the new transverse components, $B_x^{\prime}$ and $B_y^{\prime}$ (bottom row of Fig. \ref{fig:azim_experim_B}). Because the axis of rotation is parallel to the $B_y=B_y^{\prime}$ direction, this component remains unchanged. $B_x$ goes from having a symmetric pattern around the y-axis to being almost completely negative when projected onto the LOS reference frame ($B_x^{\prime}$). This shows how a magnetic field azimuth that is circularly symmetric in the original reference frame, is pointing predominantly to the observer in the LOS reference system.

\begin{figure}[t!]
\centering
\includegraphics[width=1.0\textwidth]{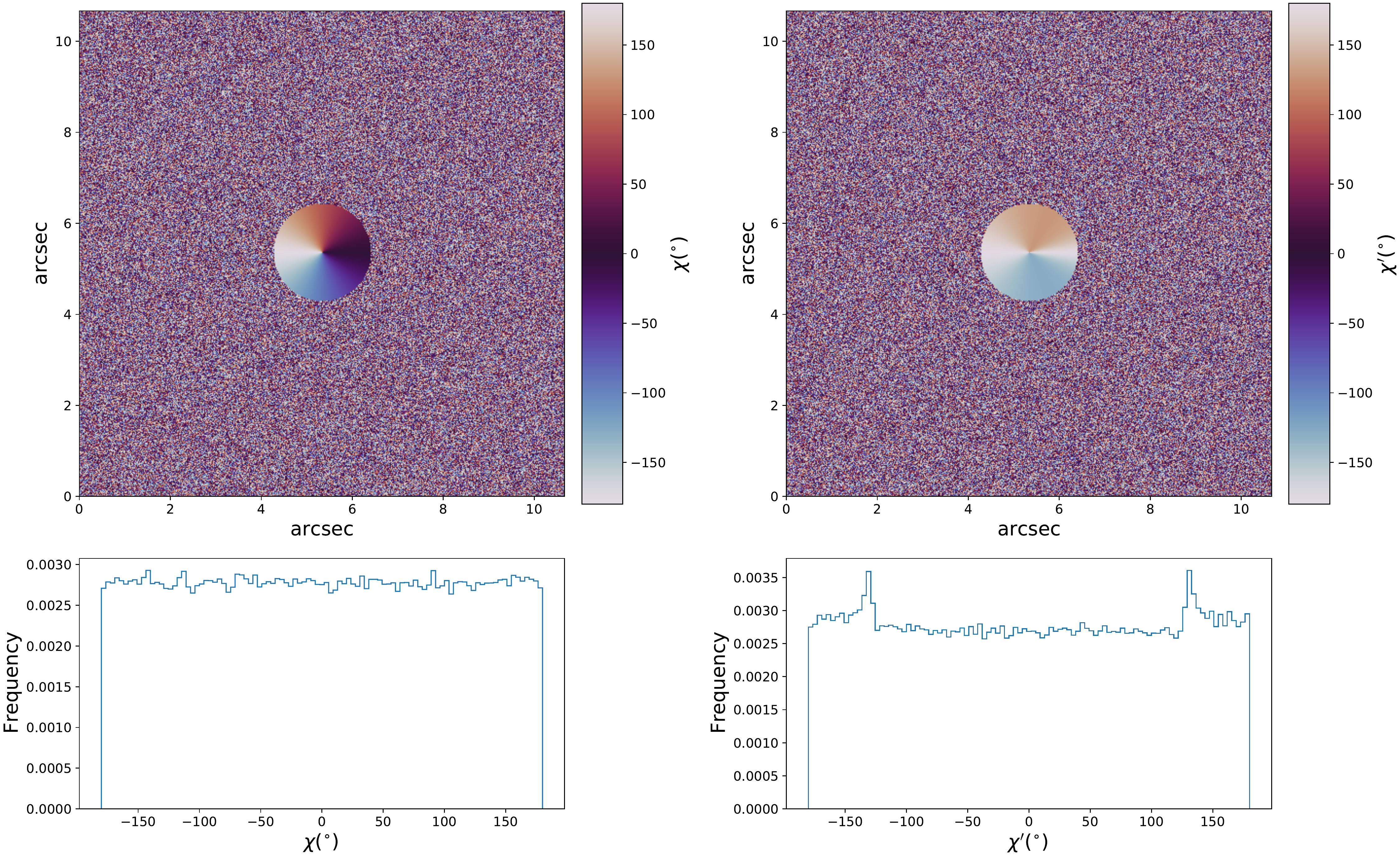}
\caption{Azimuths in the original ($\chi$) and LOS ($\chi^{\prime}$) reference frames. The azimuth images are shown in the top row and their distributions in the bottom row.   \label{fig:azim_experim_orig}}
\end{figure}

Figure \ref{fig:azim_experim_orig} shows the azimuth of the transverse magnetic component ($\chi = arctg(By/Bx)$), as well as its probability density function in these two reference frames. Outside of the magnetic patch, the azimuths are randomly oriented, and stay randomly oriented after projecting the magnetic field onto the LOS reference system (i.e. in both cases they result in uniform distributions of the azimuth). Inside the magnetic patch, however, the azimuth distribution in the LOS reference frame is heavily skewed towards the observer (i.e. along the LOS), whilst in the original reference system it has a uniform distribution across all values. This is evident in the histograms of the bottom row, where the azimuth in the LOS reference frame shows excess population around the outer edges of the distribution.

\begin{figure}[t!]
\centering
\includegraphics[width=1.0\textwidth]{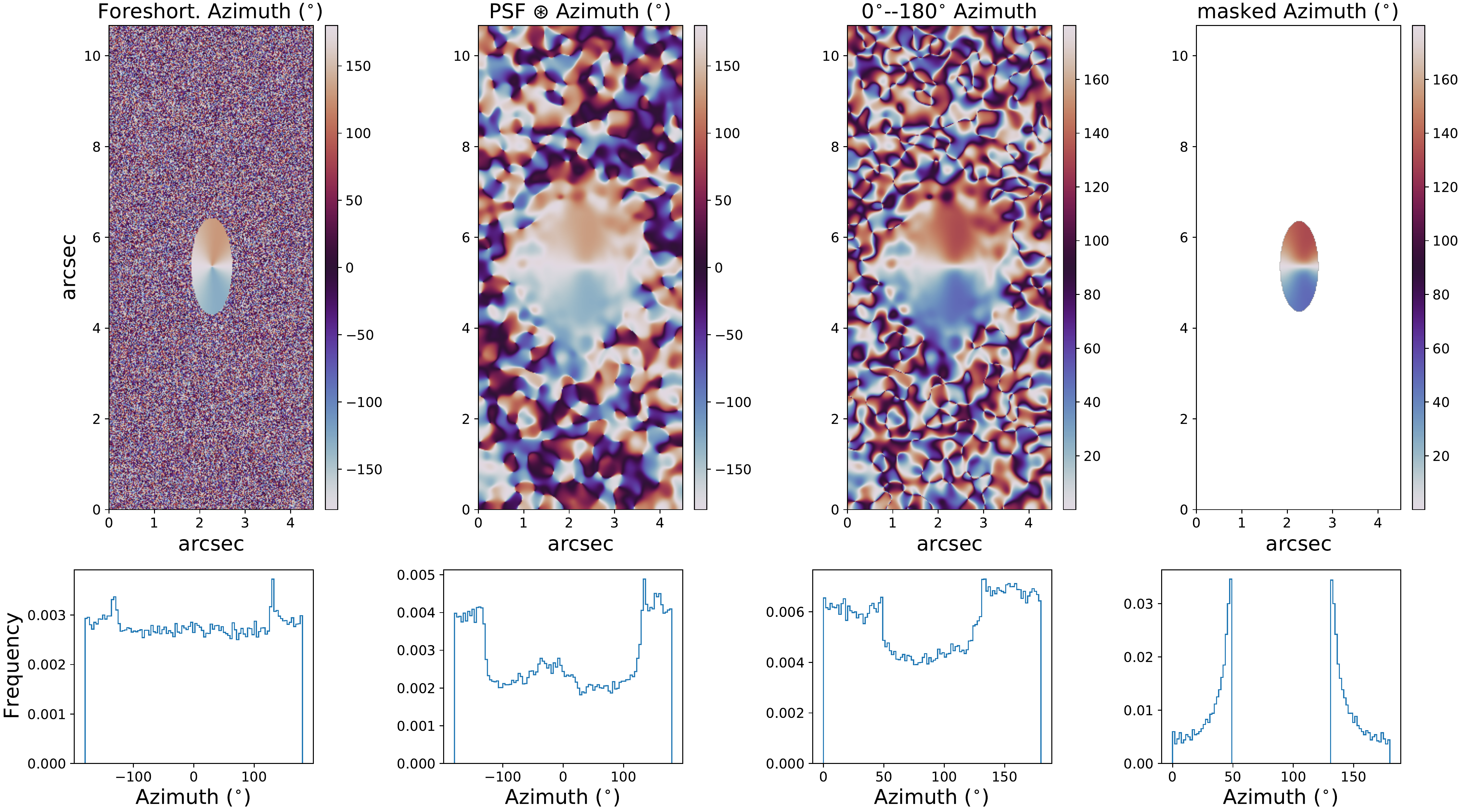}
\caption{Azimuth images (top) and distributions (bottom) for the cases where the pixels have been foreshortened (first column) a spatial degradation has been applied (second column), the azimuths have been folded between 0 and 180$^{\circ}$ (third column), and pixels with weak magnetic fields have been masked from the image (fourth column).  \label{fig:azim_experim_degraded}}
\end{figure}

When we foreshorten the deprojected magnetic field components along the LOS, the magnetic patch compresses into an ellipse. The azimuths of the foreshortened image are shown in the top left panel of Fig. \ref{fig:azim_experim_degraded}. The bottom left panel shows their corresponding PDF, which does not differ significantly from the non-foreshortened case of Fig. \ref{fig:azim_experim_orig} (right column). Next, we convolve the foreshortened vector magnetic field components with an Airy disk of 240~km radius (equivalent to a spatial resolution of ~0.32"). The resulting azimuth image (second panel in the top row of Fig. \ref{fig:azim_experim_degraded}) shows a coherent magnetic patch that looks much larger than the original one - by degrading the spatial resolution, we dilute the magnetic field and increase the apparent spatial extent that the magnetic patch occupies. The corresponding distribution function (below) presents a more pronounced skew at the edges. In the third column, we fold the azimuth between $0^{\circ}$ - $180^{\circ}$ to mimic the $180^{\circ}$ ambiguity, and in the fourth column we show the effect of restricting the analysis to pixels with strong magnetic fields. The PDF now becomes heavily skewed towards the edges of the distribution.

This experiment shows how the distribution of azimuths in an azimuthally symmetric magnetic patch changes as the magnetic field is observed in the LOS reference frame, the pixels are foreshortened, the image is spatially degraded with a PSF and the azimuths are folded between $0^{\circ}$ - $180^{\circ}$.
If we repeat this experiment for the case in which the original magnetic fields are randomly oriented, the result is a uniform distribution of azimuths at every step of the process.
While the strongly magnetized patches in the MURaM simulation are not azimuth centers, they {\em are} predominantly vertical, resulting in an overall skewed azimuth distribution when projected onto the LOS reference frame. This bias is further pronounced by spatial degradation.  
Outside of these strong patches, the magnetic fields tend to be almost randomly oriented, producing an almost uniform distribution of azimuths.


\bibliography{centeno_LWS}{}
\bibliographystyle{aasjournal}



\end{document}